\RequirePackage{lineno}
\documentclass[aps,checkin,prl,preprintnumbers,amsmath,amssymb]{revtex4}
\usepackage{graphicx}
\usepackage{dcolumn}
\usepackage{bm}
\usepackage{color}

\newcommand{\comment}[1]{}
\begin{document}
\setpagewiselinenumbers \linenumbers

\title{Magnetic ordering and structural phase transitions in strained
ultrathin SrRuO$_{3}$/SrTiO$_{3}$ superlattice }

\author{Mingqiang Gu,$^{1, 2}$ Qiyun Xie,$^1$ Xuan Shen,$^4$
\\Rubin Xie,$^1$ Jianli Wang,$^3$ Gang Tang,$^3$ Di Wu,$^4$ G. P. Zhang,$^2$
and X. S. Wu$^1$\footnote{Corresponding author: xswu@nju.edu.cn} }

\affiliation{
$^1$Laboratory of Solid State Microstructures and School of Physics, Nanjing University, Nanjing 210093, China\\
$^2$Department of Physics, Indiana State University,
Terre Haute, IN 47809 \\
$^3$Department of Physics, China University of Mining and
Technology, Xuzhou 221116, China\\
$^4$Laboratory of Solid State Microstructures and Department of Materials Science\\ and Engineering, Nanjing University, Nanjing 210093, China}

\date{\today}

\begin{abstract}
\setpagewiselinenumbers \linenumbers

Ruthenium-based perovskite systems are attractive because their
structural, electronic and magnetic properties can be systematically
engineered. SrRuO$_3$/SrTiO$_3$ superlattice, with its period
consisting of one unit cell each, is very sensitive to strain
change. Our first-principles simulations reveal that in the high
tensile strain region, it transits from a ferromagnetic (FM) metal
to an antiferromagnetic (AFM) insulator with clear tilted octahedra,
while in the low strain region, it is a ferromagnetic metal without
octahedra tilting. Detailed analyses of three spin-down
Ru-$t_\text{2g}$ orbitals just below the Fermi level reveal that the
splitting of these orbitals underlies these dramatic phase
transitions, with the rotational force constant of RuO$_6$
octahedron high up to 16 meV/Deg$^2$, 4 times larger than that of
TiO$_6$. Differently from nearly all the previous studies, these
transitions can be probed optically through the diagonal and
off-diagonal dielectric tensor elements. {\color{black}  For one
percent change in strain, our experimental spin moment change is
$-0.14\pm0.06$ $\mu_B$, quantitatively consistent with our
theoretical value of $-0.1$ $\mu_B$.}
\end{abstract}



\maketitle

Strontium ruthenate \cite{Koster2012} belongs to a broad scope of
{perovskites} and has attracted extensive attention, due to its exotic
properties. When slab thickness decreases, its itinerant
ferromagnetic (FM) phase disappears \cite{D. Toyota2005162508-3}.
This has motivated many experimental \cite{J. Xia2009140407, Y. J.
Chang2009057201, G. Herranz2008165114,
  M. Schultz2009125444} and theoretical investigations
\cite{J. M. Rondinelli2008155107, Rondinelli2010113402, P.
Mahadevan2009035106} {to search the origin of the loss of the FM
metallic phase. Another perovskite, SrTiO$_3$ (STO), is a good
insulator, and also has attractive physical properties such as
superconductivity and two-dimensional electron gas on its surface
and interfaces} \cite{Santander-Syro2011189, Li7762}. It would be a
{fascinating} idea to investigate a superlattice which consists of one
layer of ``conducting'' SrRuO$_3$ and one ``insulating'' layer of
SrTiO$_{3}$, or SRO/STO superlattice. Experimentally, thicker
superlattices have already been fabricated \cite{Herzog2010161906,
Woerner200983, M. Izumi1998227-233, M. Izumi1998651}. The Curie
temperature decreases \cite{M. Izumi1998227-233, M. Izumi1998651}
with the decrease of the period of the superlattice. The magnetic
moment of the Ru atom is suppressed, and no FM ordering was
identified in a 1/1 superlattice. {A question is thus
raised whether there is indeed any magnetic ordering established in
the SRO/STO 1/1 superlattice.}

With the advent of the state-of-the-art molecular-beam epitaxy
(MBE), pulse laser deposition (PLD) and other growth techniques, it
is now possible to fabricate interfaces with atomic sharpness. This
is particularly true for systems with a very small lattice mismatch,
such as SRO/STO. More importantly strain can be controlled through
different substrates. For instance, TiO$_{2}$ and MgO \cite{B. S.
Kwak199414865-14879} can induce an in-plane strain
 of -4\% and 7\%, respectively. Piezoelectric substrates like
PMN-PT even allow one to control the strain in real time \cite{Z.
Kutnjak2006956}. This greatly facilitates materials engineering.

In this {\it Letter}, we report that such a superlattice undergoes
an intriguing magnetic phase transition under epitaxial strain. Our
experiment first shows that in a well prepared SRO/STO superlattice
sample, ferromagnetic ordering can survive down to ultrashort period
(1/1). We can tune its magnetic properties by applying different
strains. Our first-principles calculations further reveal that the
in-plane strain can drive the system from a ferromagnetic to an
antiferromagnetic phase at a critical strain $\xi_c=5\%$. Within the
ferromagnetic phase, three structurally different phases are
identified: below 0.25\% ($\alpha$ phase), the RuO$_6$ and TiO$_6$
octahedra rotate in the opposite direction but without tilting;
between 0.25\% and 2.5\% ($\beta$ phase) tilting starts and the
rotation angles of both RuO$_6$ and TiO$_6$ are reduced; and above
2.5\% ($\gamma$ phase), both RuO$_6$ and TiO$_6$ rotate in the same
direction. To understand these dramatic changes, we carefully
examine the borderline between phases and find that the frontier
spin-down Ru-$t_\text{2g}$ orbital is mainly responsible for the
phase transition, where its occupation changes with the strain.
These phase transitions directly lead to a qualitative difference in
dielectric tensor, a signature that can be probed experimentally.

\begin{figure}
 {\includegraphics[width=1\textwidth]{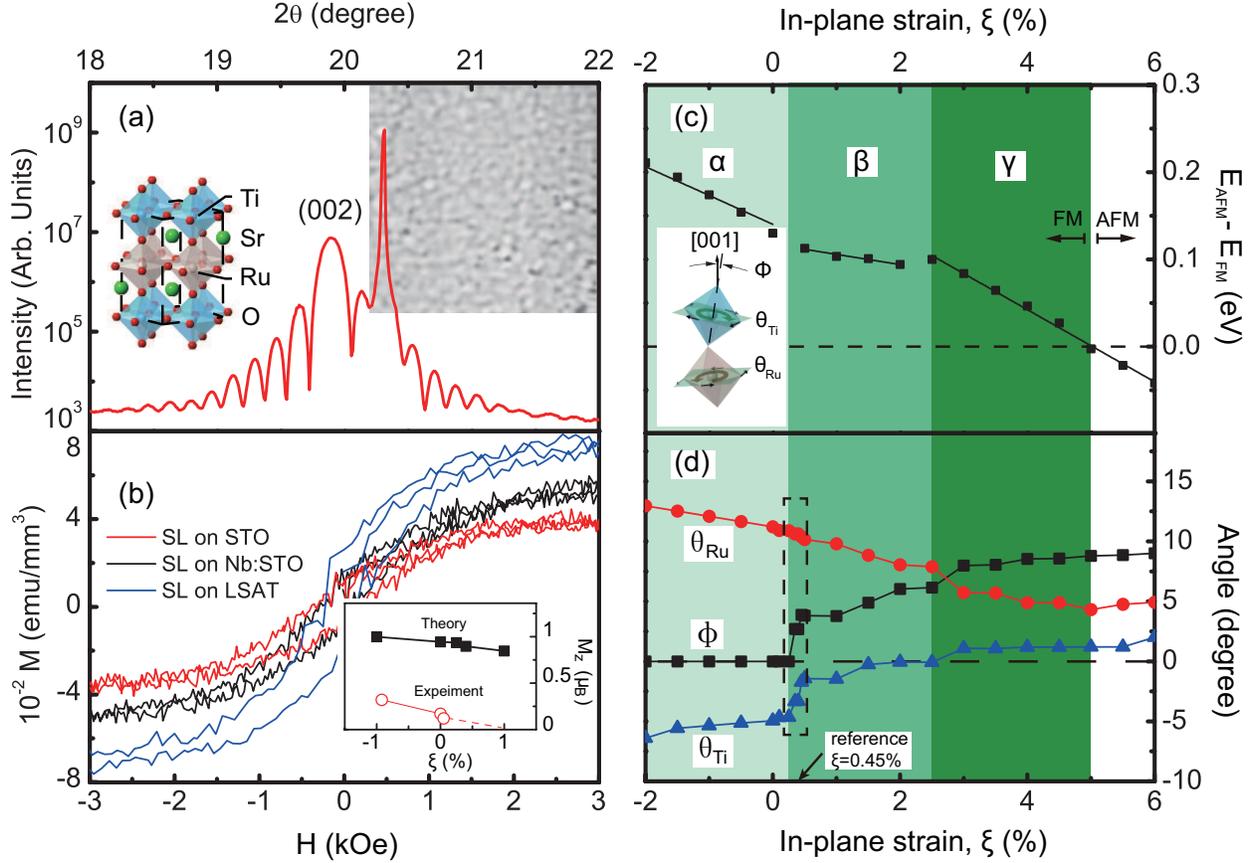}}
  \caption{\label{fig:fig1} (Color online) (a) X-ray diffraction
  pattern for the sample grown on STO substrate.
  Left inset: Superlattice structure, where
  Sr, Ru, Ti and O atoms are shown in green, brown,
  blue and red, respectively.
  Right inset: Surface AFM image for this sample.
   (b) Hysteresis loops for [SRO/STO]$_{30}$
   superlattices grown on
   STO, Nb:STO, and LSAT substrates, respectively. The inset shows
   the magnetic moment of the Ru atom as a function of strain. (Exp:
   unfilled symbols and theory: filled symbols.)
   (c) Theoretical in-plane strain
   dependence of the total energy difference between AFM
   and FM phases. Three phases
   $\alpha$, $\beta$ and $\gamma$
   in FM phase are highlighted by shaded in three different colors.
   The slopes for the three fitting lines are -0.03, -0.01,
   -0.04 eV/percent in strain, respectively.
   Inset shows the tilting angle $\phi$ and two
   rotational angles $\theta_\text{Ru}$ and $\theta_\text{Ti}$.
   (d) Optimized $\phi$, $\theta_\text{Ru}$ and $\theta_\text{Ti}$
   as a function of $\xi$. The dashed box is the region
   that is further examined in Fig. 2(a).
   }

\label{fig1}
\end{figure}

We grew [SRO/STO]$_{30}$ superlattice samples using laser MBE on
three different substrates: STO, Nb:STO and LaSrAlTaO (LSAT). {The growth of 1/1 superlattice films was monitored by reflection high energy electron diffraction (RHEED), which showed a layer-by-layer
growth mode, (see the Supplementary Materials for more details.)} Lattice mismatches in terms of STO lattice constant for
these substrates are 0\%, 0.05\% and -0.92\%, respectively. The
structures were characterized by x-ray diffraction (XRD) using the
synchrotron radiation \cite{note1} beamline BL14B1 of Shanghai
Synchrotron Radiation Facility (SSRF), shown in Fig. 1(a). The data
shows unambiguously that the superlattices are smooth and free of
any second phase. From the (002) peaks (see Fig. \ref{fig:fig1}(a)),
we estimate the out-of-plane lattice constants for these films to be
3.982 \AA, 3.978 \AA\ and 4.009 \AA, respectively, meaning that the
samples are strained according to the substrate lattice. {The Laue
oscillations of the peak due to the total crystalline film thickness
indicate a good [001] orientation of the film.} Surface atomic force
microscopy (the inset of Fig. \ref{fig:fig1}(a)) reveals that all
these samples have a smooth termination with the roughness below 1
uc.

The magnetic properties were measured at 5K by vibrating sample
magnetometer (PPMS VSM Option Release 1.2.4 Build 1). {\color{black}
Although the magnetization is suppressed compared to the bulk SRO,
clear hysteresis loops (M-H) are observed in all our samples.}
Superlattice recovers some of the lost FM ordering in SRO thin
films. To our knowledge, this has not been reported before at 1
uc-thick SRO thin films and superlattices. Though the coercive field
weakly depends on substrates, both the remanence and saturation
field strongly depend on the substrates.  The spontaneous
magnetization changes from $0.12\pm0.03$ $\mu_B$ (on Nb:STO) to
$0.30\pm0.03$ $\mu_B$ (on LSAT) per Ru. { The Ru
magnetic moment decreases as the strain increases (see the inset in
Fig. 1(b)). For one percent change in strain, the magnetic moment
change is -$0.14\pm0.06$ $\mu_B$.} Since film thickness and growing
conditions are the same for all our samples, the lattice mismatch,
or the strain, is directly responsible for the magnetic properties
change.

To understand the strain effects in this superlattice, we resort to
first-principles calculations. For bulk SRO, extensive calculations
have been performed to investigate changes in structural,
electronic, and magnetic properties \cite{J. M.
Rondinelli2008155107, A. T.
  Zayak2006094104, S. Middey2011014416, A. T. Zayak2008214410,
  H.-T. Jeng2006067002},
but very few on a superlattice \cite{Medici2011, M.
Stengel2006679-682, Verissimo-Alves2012}. We carry out
first-principles calculations on a STO/SRO superlattice (see the
inset in Fig. 1(a)) within the local spin density approximation plus
Hubbard on-site Coulomb repulsion (LSDA+U) \cite{S. L.
Dudarev19981505, note1}. Within this scheme, the magnitude of
$U_\text{eff}$ is treated as an empirical parameter, which will be
discussed in the following text and the Supplementary Materials.

The in-plane tensile strain $\xi$, defined as
$\xi=(a-a_\text{STO})/a_\text{STO}$, is applied to the superlattice.
Here $a$ is the lattice constant in use while $a_\text{STO}$ is that
of the parent compound. The in-plane strain is changed from -4.5\%
to 6\%. At each strain, the ion positions and the out-of-plane
lattice constant are fully relaxed. There are three distinctive
angles to characterize the structure change. One is the tilting
angle $\phi$ around an axis parallel to (001) plane (see the inset
in Fig. \ref{fig1}(c)). The other two angles $\theta_\text{Ru}$ and
$\theta_\text{Ti}$ denote rotations around [001] axes of the
RuO$_{6}$ and TiO$_{6}$ octahedra, respectively.

Figure \ref{fig1}(c) shows the energy difference between FM and AFM
phases as a function of strain. In a wide range of strain, the FM
alignment between the neighboring Ru atoms is energetically favored.
The magnetic moment is 2 $\mu_{B}$ per SRO formula unit, which is
consistent with the calculated ferromagnetic ground state of bulk
SRO \cite{H.-T. Jeng2006067002}. However, when the tensile strain
exceeds $\sim$5\%, the structure with {c-type} AFM phase becomes
more stable than that of FM. Therefore, a magnetic phase transition
occurs at this critical strain. One notices that the total energy
difference between the AFM and FM structures is linear with respect
to strain but with three different slopes (see the caption of Fig.
1). This suggests structural phase transitions taking place as a
consequence of strain variation.

The observation is indeed verified. Structurally, the FM region can
be subdivided into three different phases: $\xi\leqslant0.25\%$
($\alpha$ phase), $0.25\%<\xi\leqslant2.5\%$ ($\beta$ phase) and
$2.5\%<\xi\leqslant5\%$ ($\gamma$ phase) (see Figs. 1(c) and 1(d)).
Within $\alpha$ phase the neighboring octahedra rotate
counterclockwise with respect to each other, where
$\theta_\text{Ru}$ is always positive but $\theta_\text{Ti}$ is
negative (see Fig. 1(d)). There is no tilting, i.e. $\phi=0^\circ$.
The quenching of tilting renders the superlattice with a high
symmetry of $P4/mbm$, making it easy be detected (see below).
Similar observations have been reported experimentally in single
crystals \cite{A. Vailionis2008051909-3, Rondinelli2010113402,
He2010227203, Borisevich2010087204}. Once the strain exceeds
$0.25\%$, the symmetry is reduced to $P2_1/c$. The $\beta$ phase
features a tilting ($\phi\neq0^\circ$) and two octahedra TiO$_6$ and
RuO$_6$ rotating in the opposite direction (see $\theta_\text{Ti}<0$
and $\theta_\text{Ru}>0$). In the $\gamma$ phase the TiO$_{6}$
octahedron rotates in the same direction as RuO$_{6}$.

\begin{figure}
 {\includegraphics[width=1\textwidth]{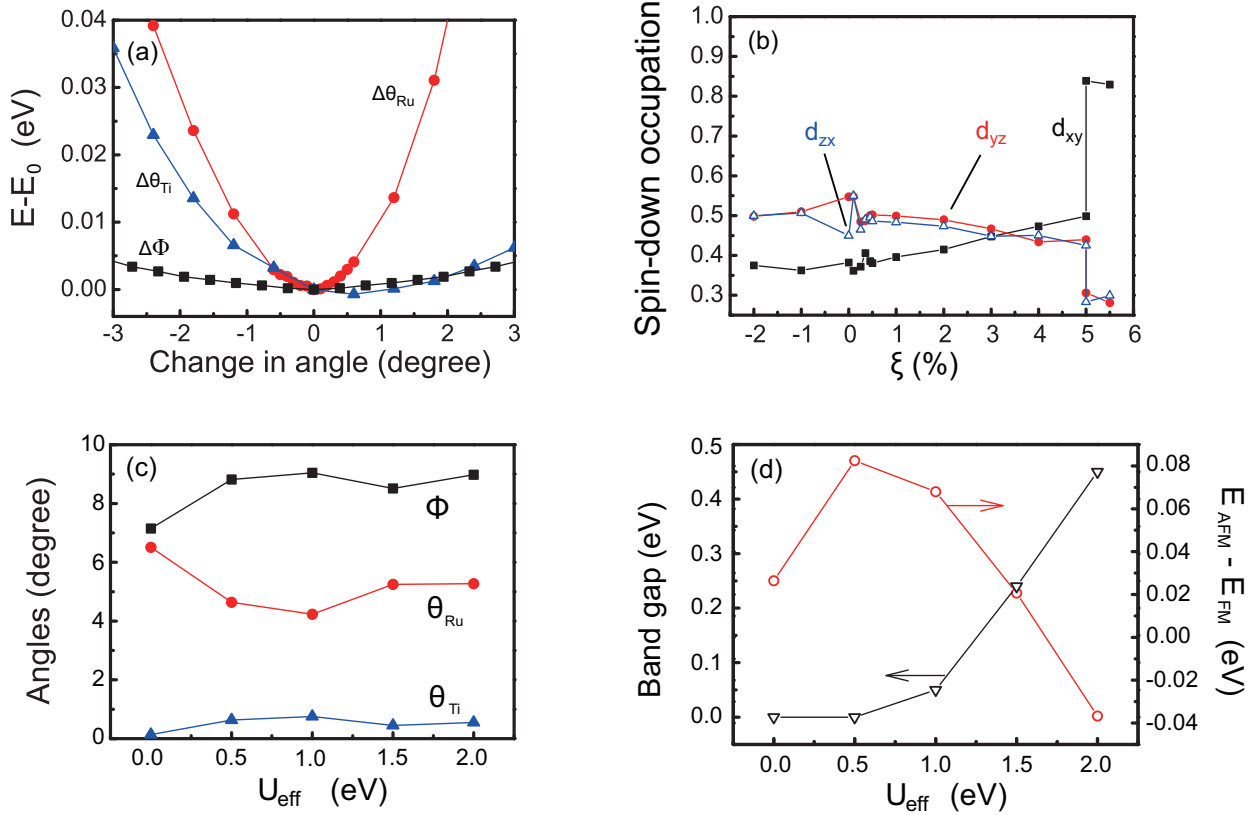}}
  \caption{\label{fig:fig2}(Color online) (a) Relative total energy
  as a function of the rotational and tilting angles.
  Here $\Delta\phi=\phi-\phi^0,
   \Delta\theta_\text{Ru}=\theta_\text{Ru}-\theta_\text{Ru}^0,
   \Delta\theta_\text{Ti}=\theta_\text{Ti}-
   \theta_\text{Ti}^0$, where $\phi^0$, $\theta_\text{Ru}^0$ and
   $\theta_\text{Ti}^0$ are their respective
   equilibrium values at the critical point
   ($\phi^0=0^\circ, \theta_\text{Ru}^0=10.9^\circ,
  \theta_\text{Ti}^0=-4.7^\circ$). $E_{0}$ is the energy of the equilibrium
  structure. $\Delta\theta_\text{Ti}$ has its lowest
  point at $+0.6^\circ$, but this is within our relaxation
  accuracy of 2 meV.
  (b) Spin-down occupation of the three Ru $t_\text{2g}$ orbitals.
  See spin-up occupation in the Supplementary Materials.
  The occupation is calculated by
  integrating the projected DOS from -10 eV to the Fermi
  level within the Wigner-Seitz radius of 1.402 \AA.
  (c) Tilting and rotation angles at $\xi=6\%$ as a function of $U_\text{eff}$.
  (d) Band gap for the AFM phase and total energy difference as a function of
      $U_\text{eff}$.}

\label{fig2}
\end{figure}

To shed light on this structural phase transition, we zoom in a
small region around $\xi=0.25\%$ (see the small dashed box in Fig.
\ref{fig1}(d)). We manually change four structural parameters
($\phi$, $\theta_\text{Ru}$, $\theta_\text{Ti}$ and $\xi$) around
their respective equilibrium values while keeping the rest
unchanged. To make a quantitative comparison, we choose the
structure at $\xi=0.45\%$ with a tilting angle $\phi=3.88^\circ$ as
the reference structure since it is near the critical point. The
energy difference curves are plotted in Fig. \ref{fig2}(a). The
energy minimum is indeed at its global minimum, since each curve has
a minimum at 0$^\circ$, with a small deviation in
$\Delta\theta_\text{Ti}$ due to our energy threshold of $\pm1$ meV
in our optimization procedure. Contributions from each angle are
very different. The rotational angle $\Delta\theta_\text{Ru}$ has
the strongest effect on the energy change among all the angles. If
we fit the energy change to a harmonic potential
$\frac{1}{2}K_{\Delta\theta_\text{Ru}}(\Delta\theta_\text{Ru})^2$,
we find $K_{\Delta\theta_\text{Ru}}=16$ meV/Deg$^2$, or 53
eV/rad$^2$ \cite{gpzhang2005}. This is much larger than that of
$K_{\Delta\phi}=1$ meV/Deg$^2$ and $K_{\Delta\theta_\text{Ti}}=4$
meV/Deg$^2$. In addition, the energy change of
$\Delta\theta_\text{Ti}$ is highly anharmonic. We expect
experimentally Raman spectra can easily distinguish them.

It is conceivable that the above strongest contribution from
$\Delta\theta_\text{Ru}$ must be associated with the electronic
structure of Ru ions. To see this, we integrate the Ru-$t_\text{2g}$
projected density of states from -10 eV to the Fermi level. Figure
2(b) shows that the occupation of these orbitals changes
substantially with strain. If we compare its change with the phase
change in Figs. 1(c) and 1(d), we find a very nice match between
them. {\color{black} Starting from the strain away from the critical
one, the occupations in $d_\text{zx}$ and $d_\text{yz}$ are similar.
Near the phase boundary of the structural phase transition their
occupations differ from each other. The $d_\text{yz}$ orbital gains
electrons while the $d_\text{zx}$ loses electrons. After tilting
sets in, the occupations of these two orbitals again become close to
each other. No other element has these characteristic changes.
Moreover, since the Fermi level is mostly contributed by the Ru-O
hybridized states, this explains why the energy change with respect
to the rotation of RuO$_6$ is most pronounced. At the FM/AFM phase
transition point, electrons are transferred from the
$d_\text{yz/zx}$ to the $d_\text{xy}$ orbital. One spin down
electron resides almost entirely in the $d_\text{xy}$ orbital. More
details are provided in the Supplementary Materials.}

We also investigate the effect of the Hubbard $U$ on the above phase
transitions. Structurally, $U_\text{eff}$ has a minor effect
\cite{J. M. Rondinelli2008155107}. Changes in the rotation angles at
strain $\xi=1\%$ and $\xi=4.5\%$ are too small to show. The largest
change of about 2 degrees in the RuO$_6$ rotation (Fig.2(c)) is
found at a highly strained case ($\xi=6\%$). But none of these
affects the above phase separation. Therefore, the $P4/mbm
\rightarrow P2_1/c$ transition is robust \cite{note1}. Figure 2(d)
shows that the band gap for the AFM phase is established for
$U_\text{eff}\ge1$ eV, and the total energy favors the AFM phase for
$U_\text{eff}>1.5$ eV. This is expected since it is well known that
the strong on-site correlation favors an AFM phase \cite{Fulde}.

{\color{black} Since $U_\text{eff}$ effectively splits and shifts
band states, the spin moment is fixed. To compare with our
experimental spin moment change, we set $U_\text{eff}$ to zero
\cite{note2} and compute the spin moment change. The results are
shown in the inset of Fig. \ref{fig:fig1}(b). For every percent
change in strain, our theoretical magnetic moment change of -0.1
$\mu_B$ agrees with our experimental value of -$0.14\pm0.06$ $\mu_B$
quantitatively.}

\comment{In order to make comparison between our experiment and
theory, we check the change in magnetic moment of the Ru atom. We
find that the experimental results can be reproduced theoretically
with a small value of $U_\text{eff}$, since large $U_\text{eff}$
always conceals the response of magnetic moment with a fully
spin-polarized state. Fig. 1(b) shows that in both our theory and
experiment, the Ru magnetic moment increases as strain decreases
within the $\alpha$ phase(see more discussion in Part C of the
Supplementary Materials). Therefore, the result from LSDA+U is
consistent with our experiment up to the qualitative level, though
slightly differs in detail. Additional experimental investigations
and theoretical development are necessary for a quantitative
comparison.}

Finally, we demonstrate that both predicted structural and magnetic
transitions are detectable optically. The structural phase
transition at $\xi=0.25\%$ breaks the mirror symmetry
($C_\text{2v}$), while the magnetic ordering transition changes the
band structure. The former leads to a dramatic difference in the
off-diagonal element of the dielectric tensor, and the latter leads
to another big difference in the diagonal elements. In other words,
we probe structural and magnetic phase transition using two
different tensor elements. Since VASP does not include the intraband
contribution, we decide to use Wien2K to compute the tensor since
both interband and intraband transitions are taken into account
\cite{note1}.

The off-diagonal element of the dielectric tensor is used to probe
the first order structural phase transition. In $\alpha$ phase,
$C_{2v}$ mirror symmetry exists and all the off-diagonal elements
are zero. When tilting sets in and the mirror symmetry is broken in
$\beta$ phase, nonzero $\varepsilon_\text{yz}$ emerges, a
manifestation of the beginning of octahedra tilting (see Fig.
\ref{fig3}(a) and more discussions in Supplementary Materials). At
the critical point the metal-to-insulator transition occurs. To
observe this transition, the diagonal element is used.  In the
metallic phase, the low energy excitation consists of a plasma
contribution, and the diagonal elements of dielectric tensor diverge
like $1/\omega$. In the insulating phase, on the other hand, only
interband transitions are left. The low energy divergence no longer
exists. Such change can be seen in Fig. \ref{fig3}(b). Thus the two
major phase transitions can be detected through a simple optical
setup.

In conclusion,  we have shown that SRO/STO superlattice preserves
its ferromagnetic ground state at ultra-short limit. Our experiment
has demonstrated that their magnetic properties are tunable through
different strains induced by different substrates. This is confirmed
in our theory. Our theory further reveals a strain-dependent phase
evolution for SRO/STO superlattice, where increase in strain can
drive the superlattice from a ferromagnetic metallic phase to an
antiferromagnetic insulating phase. There are three phases within
FM. In the $\alpha$ phase, the RuO$_6$ and TiO$_6$ octahedra do not
tilt, but in the $\beta$ and $\gamma$ phases, they do. We have shown
that the Ru-$t_\text{2g}$ orbital underlies these multiple-facet
changes, which can be detected experimentally. By examining the
effects of Hubbard U, we find that our theory with LSDA+U does
qualitatively support our experimental findings that the strain
induces changes in magnetic properties. Therefore, our findings are
significant as they reveal fascinating opportunities in the Ru-based
strongly correlated electronic systems, which are crucial for future
applications in ferroics and nano devices \cite{R. Ramesh200721-29,
J. H. Lee2010207204, Lee2010954-958}.

This work is supported by NKPBRC (2010CB923404), NNSFC (Nos. 11274153,
10974081, 10979017), and the U.S. Department of Energy under
Contract No. DE-FG02-06ER46304(MG, GPZ). The authors thank beamline BL14B1
(Shanghai Synchrotron Radiation Facility) for providing the beam
time. We are grateful to the High Performance Computing Center of
Nanjing University and the High Performance Computing Center of
China University of Mining and Technology for the award of CPU
hours. We also acknowledge part of the work as done on Indiana State
University's high-performance computers. This research used
resources of the National Energy Research Scientific Computing
Center, which is supported by the Office of Science of the U.S.
Department of Energy under Contract No. DE-AC02-05CH11231. Our
calculations used resources of the Argonne Leadership Computing
Facility at Argonne National Laboratory, which is supported by the
Office of Science of the U.S. Department of Energy under Contract
No. DE-AC02-06CH11357. MG thanks China Scholarship Council for the financial support for his exchange grogram and Indiana State University for the hospitality under the exchange program.


\begin{thebibliography}{27}

\bibitem{Koster2012}
G. Koster, L. Klein, W. Siemons, G. Rijnders, J. S. Dodge, C.-B.
Eom, D. H. A. Blank and M. R. Beasley, Rev. Mod. Phys. \textbf{84},
253 (2012).

\bibitem{D. Toyota2005162508-3} D. Toyota, I. Ohkubo, H.
Kumigashira, M. Oshima, T. Ohnishi, M. Lippmaa, M. Takizawa, A.
Fujimori, K. Ono, M. Kawasaki and H. Koinuma, Appl. Phys. Lett.
\textbf{87}, 162508 (2005).

\bibitem{J. Xia2009140407} J. Xia, W. Siemons, G. Koster, M. R.
Beasley and A. Kapitulnik, Phys. Rev. B \textbf{79}, 140407 (2009).

\bibitem{G. Herranz2008165114}
G. Herranz, V. Laukhin, F. S\'{a}nchez, P. Levy, C. Ferrater, M. V.
Garc\'{\i}a-Cuenca, M. Varela and J. Fontcuberta, Phys. Rev. B
\textbf{77}, 165114 (2008).

\bibitem{M. Schultz2009125444}
M. Schultz, S. Levy, J. W. Reiner and L. Klein, Phys. Rev. B
\textbf{79}, 125444 (2009).

\bibitem{Y. J. Chang2009057201}
Y. J. Chang, C. H. Kim, S. H. Phark, Y. S. Kim, J. Yu and T. W. Noh,
Phys. Rev. Lett. \textbf{103}, 057201 (2009).

\bibitem{P. Mahadevan2009035106}
P. Mahadevan, F. Aryasetiawan, A. Janotti and T. Sasaki, Phys. Rev.
B \textbf{80}, 035106 (2009).

\bibitem{J. M. Rondinelli2008155107}
J. M. Rondinelli, N. M. Caffrey, S. Sanvito and N. A. Spaldin, Phys.
Rev. B \textbf{78}, 155107 (2008).

\bibitem{Rondinelli2010113402}
J. M. Rondinelli and N. A. Spaldin, Phys. Rev. B \textbf{82}, 113402
(2010).

\bibitem{Santander-Syro2011189}
A. F. Santander-Syro, O. Copie, T. Kondo, F. Fortuna, S. Pailhes, R.
Weht, X. G. Qiu, F. Bertran, A. Nicolaou, A. Taleb-Ibrahimi, P. Le
Fevre, G. Herranz, M. Bibes, N. Reyren, Y. Apertet, P. Lecoeur, A.
Barthelemy and M. J. Rozenberg, Nature (London) \textbf{469}, 189
(2011).


\bibitem{Li7762}
L. Li, C. Richter, J. Mannhart and R. C. Ashoori, Nat. Phys.
\textbf{7}, 762 (2011).

\bibitem{Woerner200983}
M. Woerner, C. v. Korff Schmising, M. Bargheer, N. Zhavoronkov, I.
Vrejoiu, D. Hesse, M. Alexe and T. Elsaesser, Appl. Phys. A
\textbf{96}, 83 (2009).

\bibitem{Herzog2010161906}
M. Herzog, W. Leitenberger, R. Shayduk, R. M. v. d. Veen, C. J.
Milne, S. L. Johnson, I. Vrejoiu, M. Alexe, D. Hesse and M.
Bargheer, Appl. Phys. Lett. \textbf{96}, 161906 (2010).

\bibitem{M. Izumi1998227-233}
M. Izumi, K. Nakazawa, Y. Bando, Y. Yoneda and H. Terauchi, Solid
State Ionics \textbf{108}, 227 (1998).

\bibitem{M. Izumi1998651}
M. Izumi, K. Nakazawa and Y. Bando, J. Phys. Soc. Jpn. \textbf{67},
651 (1998).

\bibitem{B. S. Kwak199414865-14879}
B. S. Kwak, A. Erbil, J. D. Budai, M. F. Chisholm, L. A. Boatner and
B. J. Wilkens, Phys. Rev. B \textbf{49}, 14865 (1994).

\bibitem{Z. Kutnjak2006956}
Z. Kutnjak, J. Petzelt and R. Blinc, Nature (London) \textbf{441},
956 (2006).

\bibitem{note1} Please see supplementary materials for more detail.


\bibitem{A. T. Zayak2006094104}
A. T. Zayak, X. Huang, J. B. Neaton and K. M. Rabe, Phys. Rev. B
\textbf{74}, 094104 (2006).

\bibitem{A. T. Zayak2008214410}
A. T. Zayak, X. Huang, J. B. Neaton and K. M. Rabe, Phys. Rev. B
\textbf{77}, 214410 (2008).

\bibitem{S. Middey2011014416}
S. Middey, P. Mahadevan and D. D. Sarma, Phys. Rev. B \textbf{83},
014416 (2011).

\bibitem{H.-T. Jeng2006067002}
H.-T. Jeng, S.-H. Lin and C.-S. Hsue, Phys. Rev. Lett. \textbf{97},
067002 (2006).

\bibitem{Medici2011}
L. de' Medici, J. Mravlje and A. Georges, Phys. Rev. Lett.
\textbf{107}, 256401 (2011).

\bibitem{M. Stengel2006679-682}
M. Stengel and N. A. Spaldin, Nature (London) \textbf{443}, 679
(2006).

\bibitem{Verissimo-Alves2012}
M. Verissimo-Alves, P. Garc\'{\i}a-Fern\'{a}ndez, D. I. Bilc, P.
Ghosez, J. Junquera, Phys. Rev. Lett. \textbf{108}, 107003 (2012).

\bibitem{S. L. Dudarev19981505}
S. L. Dudarev, G. A. Botton, S. Y. Savrasov, C. J. Humphreys and A.
P. Sutton, Phys. Rev. B \textbf{57}, 1505 (1998).




\bibitem{He2010227203}
J. He, A. Borisevich, S. V. Kalinin, S. J. Pennycook and S. T.
Pantelides, Phys. Rev. Lett. \textbf{105}, 227203 (2010).


\bibitem{Borisevich2010087204}
A. Y. Borisevich, H. J. Chang, M. Huijben, M. P. Oxley, S. Okamoto,
M. K. Niranjan, J. D. Burton, E. Y. Tsymbal, Y. H. Chu, P. Yu, R.
Ramesh, S. V. Kalinin and S. J. Pennycook, Phys. Rev. Lett.
\textbf{105}, 087204 (2010).

\bibitem{A. Vailionis2008051909-3}
A. Vailionis, W. Siemons and G. Koster, Appl. Phys. Lett.
\textbf{93}, 051909 (2008).




\bibitem{gpzhang2005}
For a qualitative comparison, it is about 7 times larger than the
spring constant for the rotation in C$_{60}$, see G. P. Zhang, Phys.
Rev. Lett. \textbf{95}, 047401 (2005).



\bibitem{Fulde}
P. Fulde, {\it Electron Correlations in Molecules and Solids}
(Springer, Berlin, 1995)

\bibitem{note2}
The rigidly shifted band with a large $U_\text{eff}$ does not allow
to see the change in magnetic moment. Thus we set $U_\text{eff} =0$
here. See more discussion in the Supplementary Materials.

\bibitem{Lee2010954-958}
J. H. Lee, L. Fang, E. Vlahos, X. Ke, Y. W. Jung, L. F. Kourkoutis,
J.-W. Kim, P. J. Ryan, T. Heeg, M. Roeckerath, V. Goian, M.
Bernhagen, R. Uecker, P. C. Hammel, K. M. Rabe, S. Kamba, J.
Schubert, J. W. Freeland, D. A. Muller, C. J. Fennie, P. Schiffer,
V. Gopalan, E. Johnston-Halperin and D. G. Schlom, Nature (London)
\textbf{466}, 954 (2010).

\bibitem{R. Ramesh200721-29}
R. Ramesh and N. A. Spaldin, Nat. Mater. \textbf{6}, 21 (2007).

\bibitem{J. H. Lee2010207204}
J. H. Lee and K. M. Rabe, Phys. Rev. Lett. \textbf{104}, 207204
(2010).


\end{thebibliography}

\clearpage

\begin{figure}
{\includegraphics[width=1\textwidth]{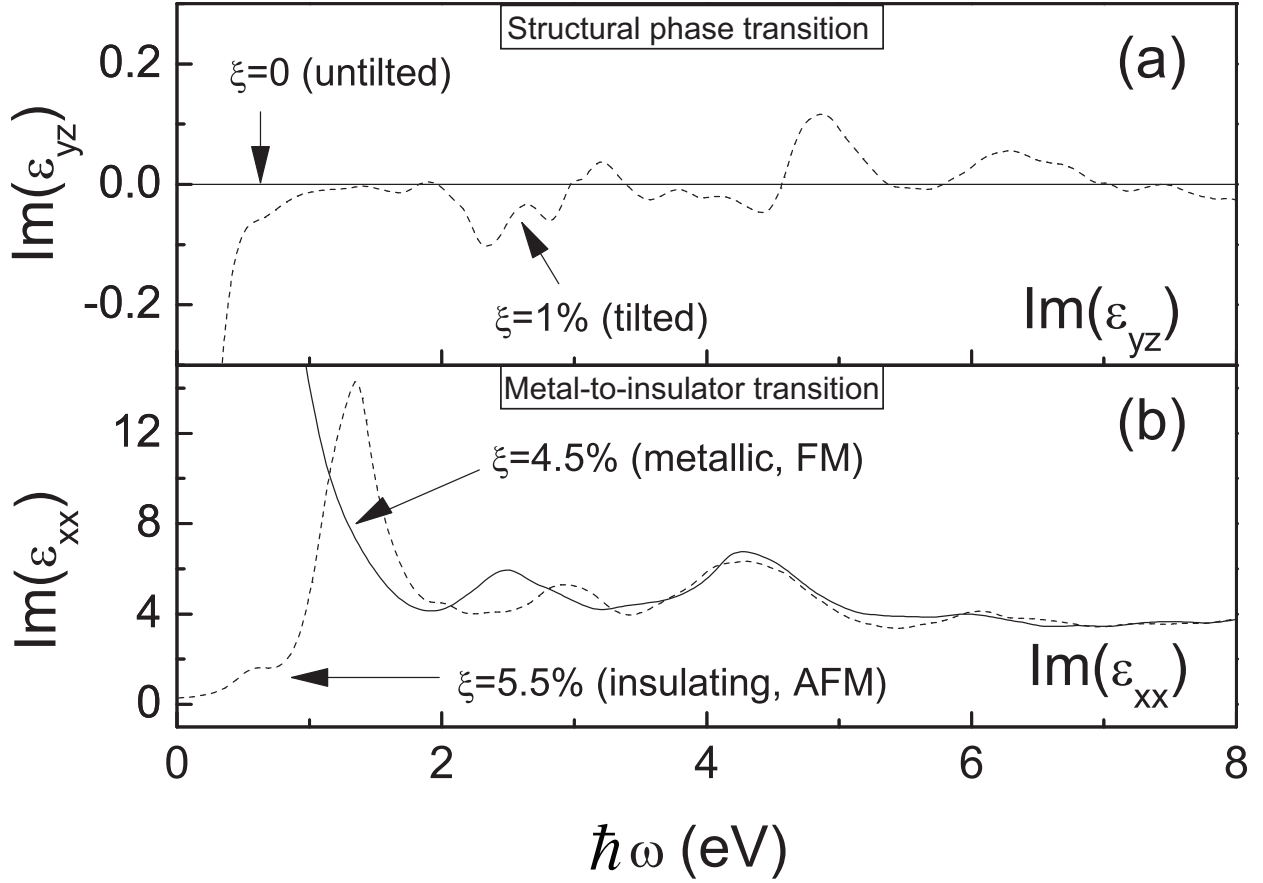}}
  \caption{\label{fig:fig3} Different optical
  tensor elements are used to monitor structural phase transition (from
  an untilted to a tilted structure), and the metal-to-insulator transition
  (from a metallic FM phase to an insulating AFM phase).
  (a) Off-diagonal element, Im($\varepsilon_\text{yz}$), as a function of photon energy
  $\hbar\omega$. The untilted structure has a null signal, while the tilted one has a
  signal.
  (b) Diagonal elements, Im($\varepsilon_\text{xx}$), as a function of $\hbar\omega$.
  The focus is on the lower energy side. In the metallic phase the Im($\varepsilon_\text{xx}$)
  diverges, while no divergence exists in the insulating phase.
} \label{fig3}
\end{figure}

\end{document}